\documentclass[twoside,english,journal]{IEEEtran}
\usepackage[T1]{fontenc}
\usepackage[latin9]{inputenc}
\usepackage{amssymb}
\usepackage{graphicx}

\makeatletter

\providecommand{\tabularnewline}{\\}

\author{J.~Jugo~\IEEEmembership{Member,~IEEE},
M.~Eguiraun,
I.~Badillo,
I.~Arredondo,
D.~Piso~\IEEEmembership{Member,~IEEE}.
\thanks{J. Jugo, M. Eguiraun and I. Badillo are with the Department of Electricity and Electronics, University of the Basque Country, Leioa, 48940, Bizkaia, SPAIN, e-mail: josu.jugo@ehu.es.}
\thanks{ I. Arredondo is ESS Bilbao, Edificio Cosimet Paseo de Landabarri 2, 48940, Leioa  (Bizkaia), Spain.}
\thanks{D. Piso is with ESS-AB, Lund, Sweden.}
}

\makeatother

\usepackage{babel}
\begin{document}

\title{Design and Performance Analysis of a Non-Standard EPICS Fast Controller}
\maketitle
\begin{abstract}
The large scientific projects present new technological challenges,
such as the distributed control over a communication network. In particular,
the middleware EPICS is the most extended communication standard in
particle accelerators. The integration of modern control architectures in
these EPICS networks is becoming common, as for example for the PXI/PXIe
and xTCA hardware alternatives. In this work, a different integration
procedure for PXIe real time controllers from National Instruments is
proposed, using LabVIEW as the design tool. This methodology is considered
and its performance is analyzed by means of a set of laboratory experiments.
This control architecture is proposed for achieving the implementation
requirements of the fast controllers, which need an important amount
of computational power and signal processing capability, with a tight
real-time demand. The present work studies the advantages and drawbacks
of this methodology and presents its comprehensive evaluation by means
of a laboratory test bench, designed for the application of systematic
tests. These tests compare the proposed fast controller performance
with a similar system implemented using an standard EPICS IOC provided
by the CODAC system. \end{abstract}

\section{Introduction}

The particle accelerators require a high variety of technologies for
a correct operation, both cutting-edge and mature. Therefore, an important
effort must be focused on the development of new technologies in order
to achieve the high demands in this field. The control tasks in a
large scientific facility are fundamental for a correct machine operation
and for ensuring the safety of the machine and the personal. The design
of those control systems is a challenging problem, which must take
into account considerations about the distributed nature of the facility
and the variety of present technologies. 

Experimental Physics and Industrial Control System (EPICS) middleware
is one of the most relevant solutions developed for the implementation
of the control solutions in this class of facilities and, in particular,
it is very usual as main control technology in particle accelerators.
This open source toolkit was initially developed for the control system
of the Advanced Photon Source (APS) in the\emph{ }Argonne National
Laboratory in collaboration with the Los Alamos National Laboratory,
\cite{APSControl}\emph{, }and is a software framework specifically
designed for the development and implementation of distributed control
systems for scientific facilities, \cite{KraimerEPICSExper}. In particular,
it is useful for the connection of large number of controllers in
a networked environment and gives monitorization, supervision and
control utilities, including feedback actions and guaranteeing soft
real-time behavior.

The evolution of EPICS over the years in different facilities reflects
its capabilities in these environments. EPICS has been evolved and adopted
in a large number of facilities due to its power and versatility and,
nowadays, it is used in many particle accelerators and other large
facilities, including telescopes, and other scientific and industrial
facilities, \cite{DiamondControl,JPARC_Control,LANSCE_Control}. Some
of the well known facilities using EPICS are the Spallation Neutron
Source in USA \cite{SNSControl} , the KEKB Collider in Japan, \cite{Akasaka2003138}
and the Diamond Light Source in United Kingdom, \cite{DiamondControl}. 

Recently, EPICS has been chosen as the key element in the control
system for the International Thermonuclear Experimental Reactor (ITER),
being the heart of the CODAC (Control, Data Access and Communication)
system, \cite{ITERStatus,CODAC}. In addition to this relevant international
center, the international project IFMIF/EVEDA, \cite{IFMIFControl}
and the European Spallation Source (ESS AB),  have chosen EPICS as
their respective control system \cite{ESS_S_CtrlBox,ControlBoxESSAB}.

On the other hand, different hardware solutions are used under EPICS
for the development of the control systems and the related diagnostics.
In this regard, it is worth stressing that one of the most relevant
characteristics required in those applications is the high reliability
and long lifetime of the selected technology, since the accelerators
are designed for an operation of decades. For this reason, the technology
involved in the implementation of any particle accelerator must be
well known and therefore, has been used along decades. Highlighted
examples are hardware elements based on the standards CAMAC \cite{CAMAC_Std},
and VME \cite{VME_Std}.

In particular, control devices based on VME have been a very relevant
elements in the last years, \cite{Toledo2003353}. Therefore, EPICS
has a wide support for hardware based in this standard. However, the
technological advance for instrumentation and control has lead to
the appearance of new standards such as PXI/PXIe and more recently,
xTCA \cite{LarsenXTCA}. Hence, following those advances different
initiatives are trying to increment the support for this kind of technologies
and its integration in EPICS. In fact, the efforts for integrating
PXI/PXIe based hardware in EPICS networks is being positively fueled
by the impulse received  in the last years.

In this sense, the development of standardized solutions for the ITER
project using PXI/PXIe platforms is a remarkable fact. The CODAC distribution
includes the device and driver support for PXI/PXIe based hardware,
in order to develop standard EPICS IOCs, \cite{CODAC}. In particular,
there is an strong interest for providing support for the National
Instruments devices which include FPGA cards for data acquisition
and control applications. The deployment of the integration of a standard
EPICS IOC for a NI PXI/PXIe based controller is described in \cite{Ruiz2012}.
An additional proposal in \cite{Ruiz2012}, similar to the one presented
in the work, is based on the implementation of an EPICS server in
a PXI/PXIe controller under LabVIEW RT. Finally, the use of a LabVIEW
based EPICS CA server, which is still under development, is proposed
in \cite{Zhukov2009}. 

\medskip{}

The main objective of this work is the description of an architecture
appropriate for the integration of NI PXI/PXIe platforms in EPICS
networks, extending and complementing the results presented in \cite{IEEE-TNS}.
The use of this hardware is oriented to the implementation of the
so called fast controllers, \cite{ITER_FastController}, which in
the current case are based on the use of FPGA cards and real time
controllers. The graphical design tool LabVIEW is used as the development
tool for the development of the control application. This architecture
has been successfully applied in a real system, specifically in the
control system for the ISHN ion source developed in ESS Bilbao, \cite{IEEE-TNS}
as well as in \cite{ISHP_RT14}.

In addition to the details of the proposed architecture and its implementation,
a test bench developed for the validation of the proposal is presented.
The results obtained from the preliminary tests are also described.
Moreover, several laboratory tests have been designed and implemented
in order to compare the reliability of the proposed system and compare
to a similar hardware which implements a standard EPICS IOC, based
on the support provided by CODAC. The initial results show that both
systems provide a comparable performance.

In this case, the controller is implemented using LabVIEW-RT and its
integration in EPICS networks is based on the use of the LabVIEW EPICS
server by means of the Data logging and Supervisory Control module
(DSC), with an additional interface to EPICS by means of a new external
IOC. The resulting system is equivalent to a standard EPICS IOC, overcoming
the limitations present in the LabVIEW EPICS server. This approach
has several advantages derived from an easier and flexible implementation
thanks to the use of the LabVIEW development environment and the availability
of commercial drivers. The use of LabVIEW eases and speeds up the
development of control structures, avoids the hardware dependent developing
costs and offers a very high compatibility with a large variety of
hardware devices used in control and data acquisition systems. These
characteristics make the system easily reconfigurable, with a versatile
EPICS integration.

On the other hand, several potential drawbacks arise, being the most
import one the reliability, which must be assured for those schemes,
since its usage in scientific facilities is not extended. In addition,
the control of the internal details of the software relays on external
companies. Anyway, the performance of this non-standard EPICS implementation
is analyzed, and satisfactory results have been obtained.

In addition, the proposed system is compared with a similar hardware,
which implements an EPICS IOC controller, developed using the tools
provided by the CODAC distribution. This comparison is based on a
test bench which simulates a real control system environment, following
the scheme proposed in \cite{Badillo2012}. This system emulates real
schemes present in scientific facilities and in particular, a particle
accelerator. A large number of process variables are considered, including
digital and analog IO signals, monitors and data processing. The work
describes the implementation of the test bench and analyzes the results
comparing both architectures. The study is focused on the reliability
of the configuration, facing the proposed solution and the EPICS standard
methodology. 

This work is organized as follows. First, the motivation of this work
is presented in Section 2. Next, Section \ref{sec:Integraci=0000F3n-de-controladores}
describes the main characteristics of the proposed architecture for
PXI based fast controllers and its integration in EPICS networks.
In Section \ref{sec:Tests-y-pruebas}, the laboratory test bench implemented
for testing the proposed solution and the obtained results are presented.
Finally, the conclusions and future work end this paper.

\section{Reliability of COTS control systems for particle accelerators}

The control system of a particle accelerator must fulfill multiple
technological challenges, therefore an efficient design architecture
for the control system must be considered, taking into account all
of the requirements for the whole facility. In particular, some applications
needs stringent requirements: a fast response, precise real-time measurements,
fast data acquisition, etc. In those cases, an advanced hardware is
needed to cope with the requirements, and to get the so called fast
controller, a high performance controller. In order to get the required
characteristics, the use of the most advanced and commercial available
control hardware is a valuable choice, taking into account the associated
costs. In fact, the usage of COTS hardware usually requires a lower
cost than home-made \emph{ad hoc }designs, which involves a time and
budget consuming design and development process\emph{.}

An architecture which is gaining relevance in modern accelerators
is the use of control chassis based on PXI/PXIe platforms. Furthermore,
National Instruments (NI) company provides this technology and supports a
large variety of DAQ and FPGA cards. Among other reasons, this fact
has push ITER to select NI company as one of its hardware providers.

However, in the case of being EPICS the main control element in the
facility, the key issue for the use of the aforementioned technology
provided by NI relies in its integration in EPICS networks. This is
the case when designing fast controllers based on NI PXI/PXIe technology.
Due to the interest aroused, NI provides tools integrated in its graphical
design environment LabVIEW, which allow the creation of EPICS controllers
using PXI/PXIe based high performance hardware. However, the currently
available solution has limited functionalities regarding EPICS environment
and the integration in EPICS networks is basic. The resulting functionalities
can be insufficient, depending on the specific necessities for a particular
application. 

On the other hand, LabVIEW is a versatile and powerful design tool
which allows an easy implementation of control and data acquisition
systems. The use of commercial and well tested controllers and cards
and the design of custom monitorization tools in an easy way are also
valuable characteristics. Those features make LabVIEW an interesting
design tool for specific control systems in particle accelerators.

The main advantage of this approach is that the development process
is easier and faster compared to other alternatives (usage of embedded
C within a real-time Linux for example). Control and monitorization
elements are more easily deployed when programming is done using a
top-level language such as LabVIEW. In addition, a large variety of
robust hardware is available.

Nevertheless, LabVIEW based development is uncommon in scientific
facilities. It is used mostly for laboratory developments and diagnostics.
In SNS \cite{long:02B722}, LabVIEW is used mainly for the development
of beam diagnostics. A similar path is followed in ISIS \cite{diagnostics-isis},
where the hardware deployed using LabVIEW is used mainly for beam
diagnostics and it is not integrated in the main control system. On
the contrary, in ESS Bilbao the use of LabVIEW is more extensive,
including the development of the control systems for several subsystems
of the accelerator, \cite{IEEE-TNS}. In any case, the reliability
related issues are not fully tested.

As a consequence of the previously mentioned drawbacks, the use of
COTS controllers designed with LabVIEW are not extensively tested
in current accelerators and the characteristics needed in this kind
of facilities are not conveniently assured. Summarizing, the following
characteristics must be fulfilled by any element present in the control
system of a particle accelerator:
\begin{itemize}
\item Reliability: a device including a controller developed using LabVIEW
must be carefully tested in realistic experimental environments. Only
if the concluding results are positive enough, the device is valid
to be used in the facility.
\item Standardization: due to the characteristics and limitations and the
harsh environment the installation and machine maintenance are complicated
and resource consuming tasks. For this reason, the use of standard
devices is fundamental, based on hardware, procedures and software
known by all the developers and operators. The standardization also
facilitates the evaluation and fulfillment of the required design
and operational rules.
\item Software reuse: the use of well known and documented code, enhances
software reusability and maintenance. In addition, tracing faults
and debugging can be performed more easily. 
\end{itemize}
The last two issues can be solved using a modular design architecture
and standard hardware, including well structured software libraries
and version control tools. Furthermore, the documentation is completely
necessary for achieving design goals.

Finally, the reliability of the devices can be assured under an exhaustive
test plan. In fact, this is one of the objectives in this work, thus,
the experimental test of the proposed methodology for the integration
in EPICS networks of fast controllers based on modern and commercial
equipment.

\section{\label{sec:Integraci=0000F3n-de-controladores}Integration of the
PXI based fast controllers in EPICS networks}

The use of chassis based on the PXI/PXIe bus can be considered a good
option for modern accelerators. In particular, the National Instruments
company provides this technology including FPGA and data acquisition
cards. In addition, NI has developed tools allowing the definition
of EPICS servers with LabVIEW, but with limited functionalities. The
advantage of this solution is the use of LabVIEW as design tool and
very versatile and powerful hardware, with an easy implementation.
Also, the solution includes already tested commercial drivers, and
the design of monitorization elements is is performed in a simple
way.

The EPICS solution provided by LabVIEW for the PXI/PXIe controllers
only implements a limited EPICS server by means of the Datalogging
and Supervisory Control module (DSC). Thus, it is not a complete EPICS
environment and, in many cases, it is not enough for control purposes.
The lack of features such as alarms and a full EPICS record support
is very limiting.

In consequence, some issues must be solved in order to use PXI/PXIe
fast controllers in facilities using EPICS control network: 
\begin{itemize}
\item The EPICS system provided by NI is limited and its integration is
not full, which could lead to unmet requirements depending on the
specific necessities. 
\item This kind of solutions using LabVIEW in EPICS networks is not usual
in large facilities. As result, the solutions based on LabVIEW have
not been tested adequately for being used as control elements in particle
accelerators 
\end{itemize}
The aim of the proposed solution is to deal with these problems.

\subsection{Proposed methodology}

The proposed solution to overcome the previous issues is based on
the use of the EPICS Server provided by the LabVIEW DSC module, filling
the gap with the development of a new interface in order to suit the
needs of the project. 

Initially, the primary interface between the local fast controller
and the EPICS control network is based on the tools provided by National
Instruments in the LabVIEW DSC module, and runs on the real time system
in the PXI controller. This module acts as a plug-in to the Shared
Variable Engine and functions as the link between shared variables
and the EPICS network. Shared variables are bound to an EPICS Process
Variable while the I/O server handles updates to the PVs. The I/O
Server then publishes the PVs to the network using the Channel Access
Protocol.

This interface is not sufficient, since the library provided by NI
lacks some important features (record structure, alarming, etc). and
must be improved. 

In order to solve this problem, an enhanced architecture is proposed

. It is based on a new IOC implementing a communication gateway between
both systems, the standard IOC and the LabVIEW/EPICS server in the
PXI controller. The main function of this gateway is to redirect and
complete the information from the LabVIEW/EPICS server. Hence, a full
EPICS IOC is obtained which behaves as a single system with the features
of the fast controller and with an EPICS based communication channel.

Accordingly, the most remarkable characteristics that have been added
are the following ones:
\begin{itemize}
\item Data: Tools for translating the values of the variables of the system
to EPICS are provided. The system allows the addition of information
regarding alarms, security, and so on.
\item Record structure: The specific record structures are mapped between
server and IOC. The aim of this mapping is to allow an automatic integration
and translation of variables through the proposed gateway. 
\item Timestamp: Each PV holds the time instant corresponding to its acquisition. 
\item Data archiving: It is possible to store data with EPICS archiving
tools, this facilitates the use of HyperArchiver solution. 
\item EPICS integration: Since there is a full IOC, the full set of EPICS
features is available.
\end{itemize}
\begin{figure*}
\begin{centering}
\includegraphics[width=1.8\columnwidth]{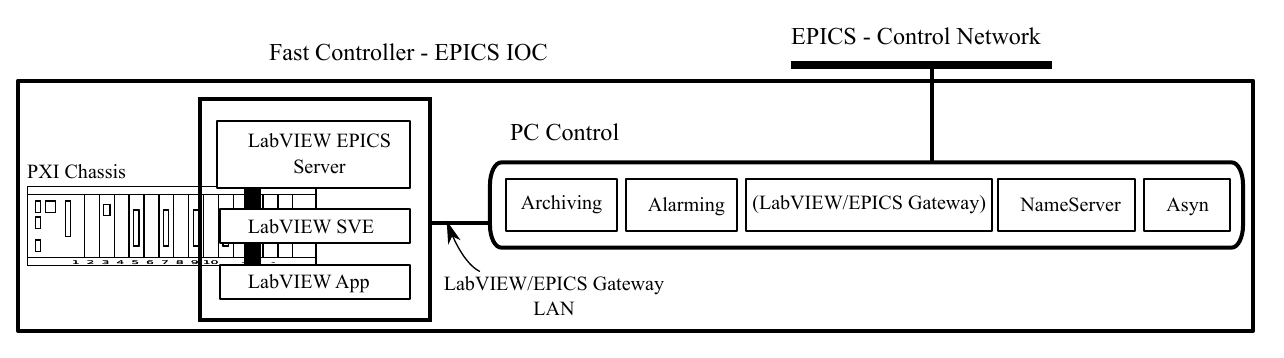}
\par\end{centering}

\protect\caption{Integration of fast controllers based on LabVIEW RT in EPICS networks.
A full IOC gets the PVs published by the EPICS server implemented
in the PXI embedded controller and adds the required functionalities\label{fig:EPICS_ISHN-1}.}
\end{figure*}

The main idea for a full integration of the fast controllers in EPICS
networks is shown in the Figure \ref{fig:EPICS_ISHN-1}. There is
an EPICS server in the LabVIEW RT system, related with a Shared Variable
Engine (SVE), which publishes the required PVs in the network. A full
IOC is connected with this server for retrieving PV data, it adds
the needed information and fields, as the information related with
the alarms, and publishes again the new PVs with a prefix identifying
them from the original ones (\emph{``NIOC:''}). Therefore, the rest
of system and applications in the network can use the new published
PVs, obtaining information related to the RT application in the fast
controller but with all the required services (data archiving, alarm
handler, monitorization, etc). The resulting control system is from
the point of view of the EPICS network, equivalent to a standard EPICS
IOC and, thus, the fast controllers is integrated in the EPICS network
in a transparent manner.

The rest of the applications on the network will use those complete
PVs in their operation (data archiving, alarming system, operator
panel, etc.) provided by the new IOC. The resulting control box can
be seen from an EPICS network as an equivalent EPICS IOC, and the
final design must be transparent from the point of view of such an
EPICS network. This approach also facilitates updates and replacement
of PXI hardware without changing the overall system characteristics. 

From the point of view of hardware requirements, an external computer
is needed for hosting the IOC which implements the gateway. This,
usually, does not present any special drawback since an standard control
cabinet has enough space for storing an embedded computer. In addition,
the cost is not increased in a dramatic way.

The following steps summarize the main rules that have to be fulfilled
for implementing a fast controller in the present methodology. The
guidelines are split into two main stages: 
\begin{enumerate}
\item Definition and programming of the record structures for the communication
gateway. In this sense, IOC records must be defined to be compatible
with the data that LabVIEW publishes. For example, a Double type could
be mapped to an analog input/output record and a DBR\_FLOAT\_EPICS
data type, a Boolean corresponds to a binary record and DBR\_ENUM,
etc. This stage is only done once, unless new functionalities are
required. 
\item Controller design

\begin{enumerate}
\item Design the control and data acquisition application for the FPGA and
the real-time (using LabVIEW design tool). 
\item Define which LabVIEW shared variables will be accessible through EPICS,
configure and deploy the required LabVIEW EPICS server in the embedded
controller. A custom program generates automatically an EPICS database
with the correct data types and links, it only requires the PV list.
\item User access control, alarm information and other functionalities are
added to each record. This step is strongly dependent on the application
objectives and decisions of the developer.
\item Deploy the IOC. Those PVs will be accessible for read and write access
in the EPICS network.
\end{enumerate}
\end{enumerate}
The addition and modification of existing features is performed following
the second step. If the improvement requires new data types or new
functionalities, new records should be added corresponding with the
new functionalities, following the rules of the first stage. This
approach for the integration of fast controllers into EPICS has been
successfully applied to the control system of the ISHN ion source
\cite{IEEE-TNS}.

In conclusion, the strategy allows the use of LabVIEW as design tool
for fast controllers, and the drivers developed by National Instruments
for their cards. However, the resulting system has the characteristics
of a full IOC controller.

This scheme allows variations and alternative architectures, such
as an hypervisor based dual boot system, \cite{HypervisorPXI}.

\section{\label{sec:Tests-y-pruebas}Performance analysis and validation tests
of the proposed methodology}

The proposed solution for the definition of the fast controllers designed
with LabVIEW and integrated in EPICS networks, has some advantages,
specially, a fast, flexible and easy development. However, the associated
risks must also be considered, specially related to the fulfillment
of the requirements requested to the hardware in a particle accelerator,
as has been mentioned in Section 2. In consequence, those risks must
be characterized and bounded to obtain similar results compared with
other technical solutions which are used nowadays in particle accelerators,
in order to be able to use the proposed solution in real environments.

\subsection{Laboratory test bench\label{sub:Banco-de-pruebas}}

The test of new control elements is not possible in a real particle
accelerator, since the real environment is not usually accessible
for those testing purposes. In addition, the real facilities are not
flexible and new devices and equipment are not easily introduced into
the system. For these reasons, a laboratory test bench has been designed
and implemented, with the aim of replicating the characteristics of
a control system in a particle accelerator, with the appropriate scaling
in its dimensions. This strategy has been followed to implement the
proposed fast controller and to compare this alternative with another
implementation based on the standard EPICS strategy. The latter has
been implemented using the tools provided by the CODAC system. In
any case, in both cases similar hardware solutions have been considered,
in order to get a realistic comparison.

\begin{figure}
\centering{}\includegraphics[width=0.9\columnwidth]{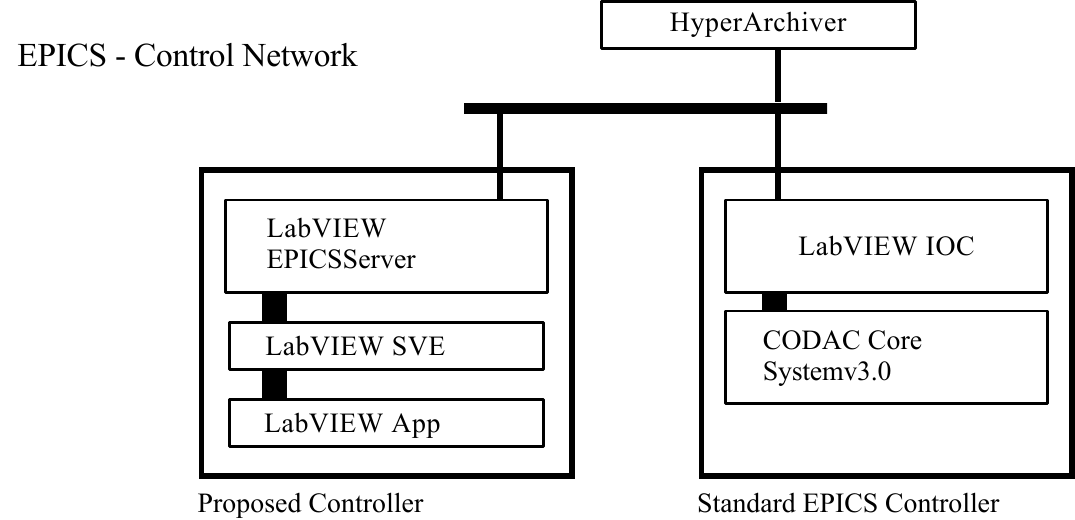}\protect\caption{\label{fig:esquemabanco}Basic scheme of the networked distributed
system for the comparison of the proposed fast controller and the
standard EPICS solution.}
\end{figure}

Therefore, the test bench includes two parallel systems which implement
two fast controllers to be tested in a long-term experiment. Both
systems implement the same functionalities and the typical actions
which are habitual in a controller for a subsystem of a particle accelerator,
which mainly include data acquisition, feedback control actions and
finite state machines. Hence, the first implementation is based on
an embedded controller under LabVIEW Real Time, publishing PVs with
the method proposed in this work, see Section 3. The second implementation
uses an embedded controller under Linux by means of a standard EPICS
IOC, \cite{Badillo2012}. This second implementation is equivalent
to the solution used in ITER, based in the CODAC system. The basic
scheme of the test bench is shown in the Figure \ref{fig:esquemabanco}.
In addition, the main characteristics of both implementations is shown
in the table \ref{tab:caracteristicas hard/soft de los teststands}.

\begin{table}
\centering{}%
\begin{tabular}{|c|c|}
\hline 
\textbf{\textit{Solution under LabVIEW RT}} & \textbf{\textit{Solution using standard EPICS}}\tabularnewline
\hline 
\hline 
Chassis PXIe-1082  & Chassis PXI-1031\tabularnewline
\hline 
Controller PXIe-8108 & ControllerPXI-8106\tabularnewline
\hline 
2.53 GHz Intel Core 2 Duo & 2.16 GHz Intel Core 2 Duo\tabularnewline
\hline 
1 GB RAM & 512 MB RAM\tabularnewline
\hline 
LabVIEW RT & Scientific Linux 6.1\tabularnewline
\hline 
LabVIEW 2011 & EPICS R3.14.12.2\tabularnewline
\hline 
\end{tabular}\protect\caption{Main characteristics of the hardware and software used in both experimental
implementations under comparison \label{tab:caracteristicas hard/soft de los teststands}.}
\end{table}

The solution based on a standard EPICS IOC is the reference for the
experimental comparison, since it is a well known solution by the
community. The IOC runs in a NI PXI-8106 embedded controller enclosed
in a PXI-1031 chassis under Scientific Linux 6.1 OS. The input/output
actions are performed using a PXI-6259 card. The drivers for this
card can be obtained from the public version of the ITER CODAC Core
System v3.0 \cite{CODAC}. The system includes a hardware-in-the-loop
(HIL) simulation of a plant to be controlled, by mean of a NI PXI-7833R
multifunction card, which includes a Virtex II FPGA. In addition,
a signal generator adds to the system the simulation of several analog
signals.

On the other hand, the proposed novel implementation to be validated
is based on a PXIe-8108 embedded controller in a chassis PXIe-1082.
In this case, the system runs under LabVIEW Real-Time OS. All of the
control actions and the input/output signals are defined in a program
developed using LabVIEW. In addition, this program implements a limited
EPICS Server, which is used to later create a full IOC and to publish
the control variables in the EPICS network. The details about this
implementation has been presented in Section 3.

\begin{figure}
\centering{}\includegraphics{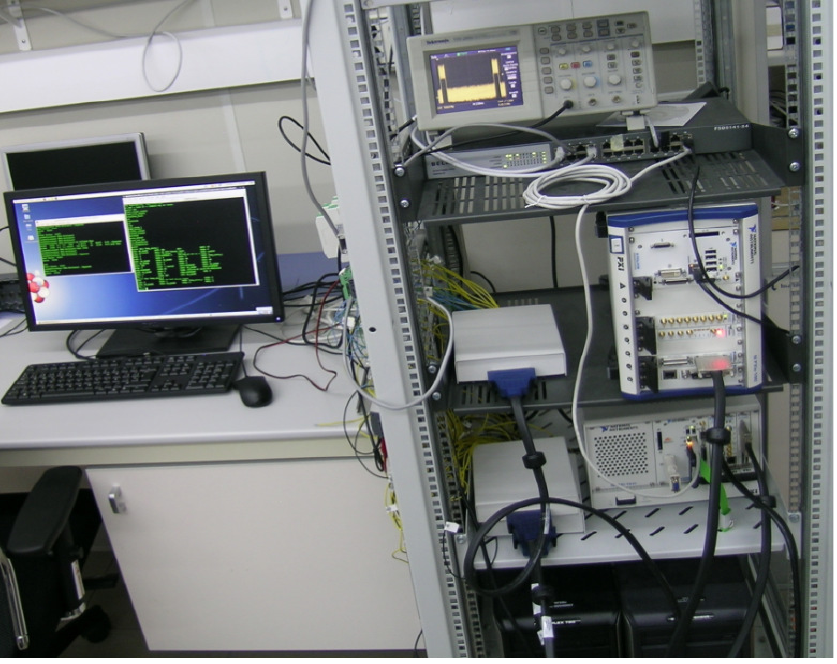}\protect\caption{\label{fig:testbench}Picture of the laboratory test bench showing
both fast controller approaches.}
\end{figure}

The test bench uses a conventional Ethernet local network, which at
the end defines the EPICS network. The overall laboratory test bench
is shown in Figure \ref{fig:testbench}.

\subsection*{Reliability test results}

\begin{figure}
\begin{centering}
\includegraphics[width=1\columnwidth]{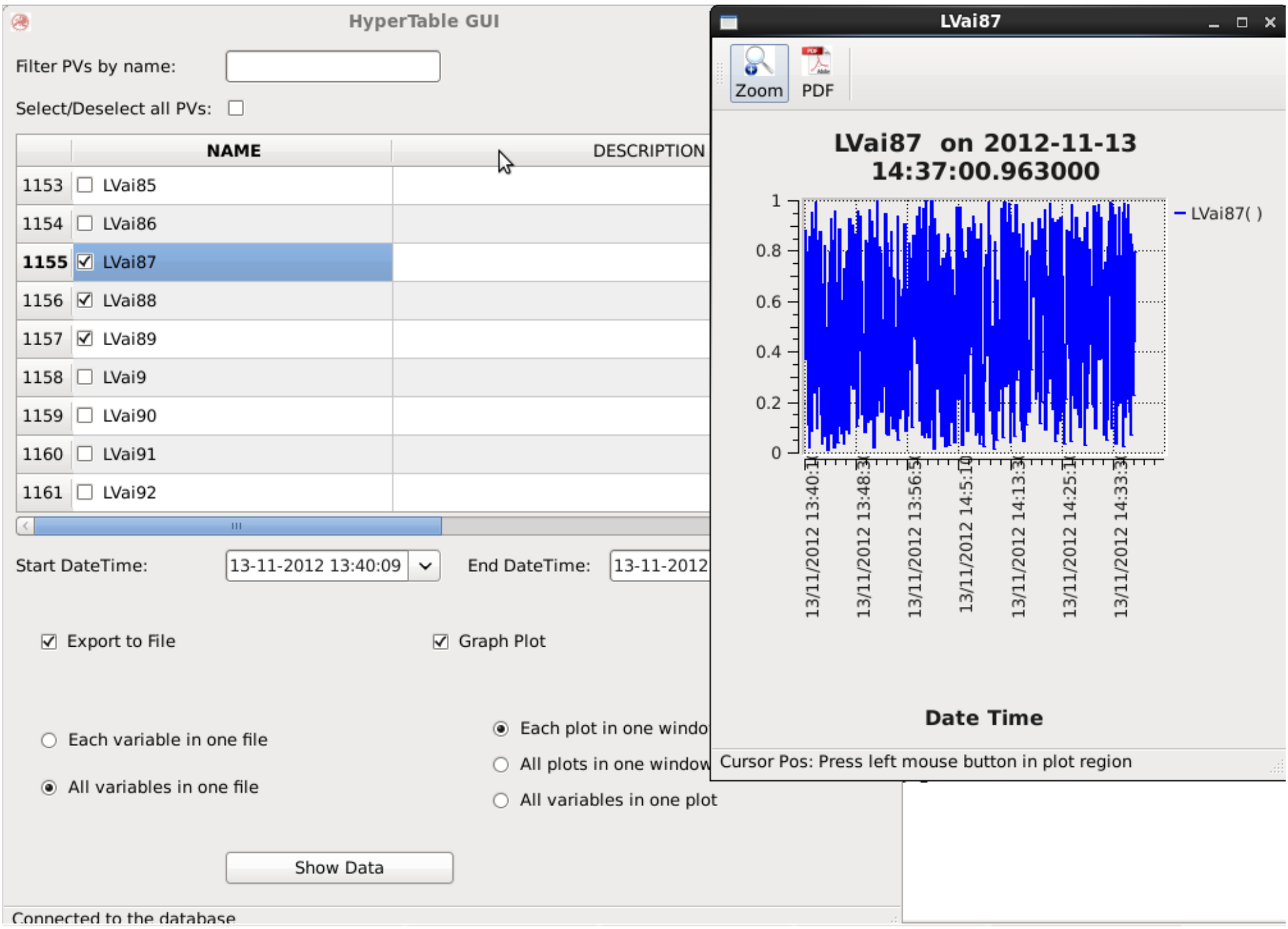}
\par\end{centering}

\protect\caption{Graphical tool developed in ESSB using pyQT for the management of
the EPICS HyperArchiver.\label{fig:HAGUI-2}}
\end{figure}

All of the results obtained with both systems are analyzed exhaustively
and archived using a dedicated database. Therefore, the reliability
of both system can be studied and compared. In this experiment, a
novel archiving system, called HyperArchiver, has been used. This
archiving system has been developed in ESS Bilbao in collaboration
with the Instituto Nazionale di Fisica Nucleare (INFN) research center
in Legnaro \cite{ESSBHyperarchiver,MariaHyperarchiver}. HyperArchiver
allows the use of large data tables with high performance thanks to
the usage of Hypertable database and it is considered scalable and
reliable. In addition, a pyQT graphical tool has been developed to
visualize the archived data (Figure \ref{fig:HAGUI-2}).

The main parameters under consideration in order to perform comparison
analysis are related to the reliability and the repeatability. The
tests carried out use 1960 records which have been implemented equally
in the both systems to compare, and then, the total number of defined
process variables is 9600, similar to the number of PVs presented
in \cite{NSLSPerformance}. Most of the PVs are processed periodically
at 1 or 5 seconds. On the contrary, a set of 760 records are event
processed. In the data archiving process, as soon as a PV is processed,\emph{
}it is written in a Hyperarchiver buffer and the resulting data sets
are batch processed each ten seconds, storing the data in the database.

\begin{figure}
\centering{}\includegraphics{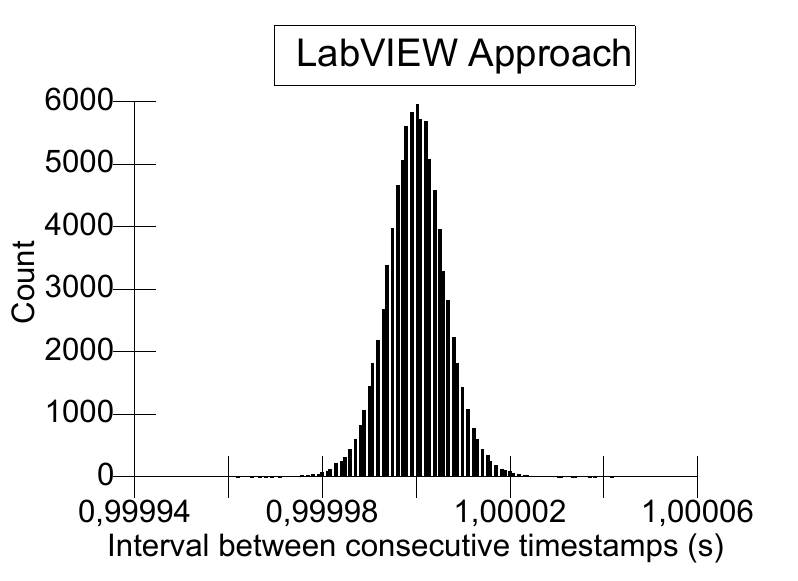}\\
\includegraphics{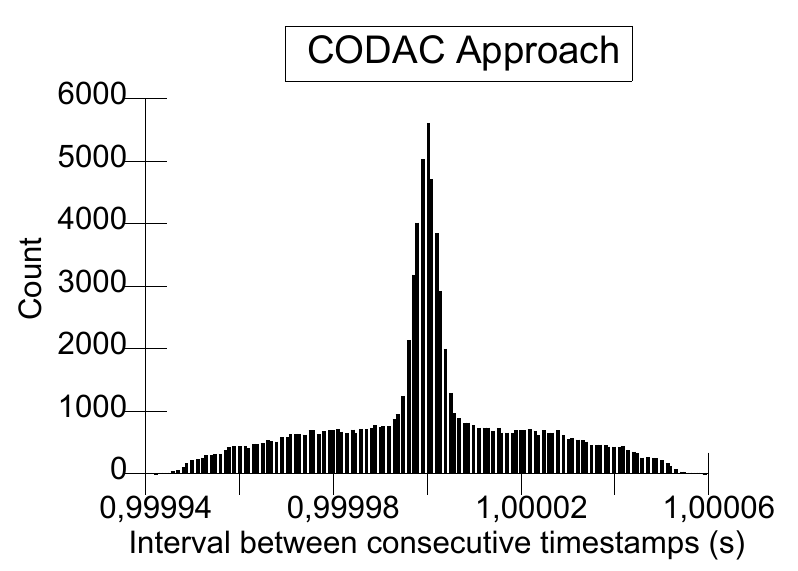}\protect\caption{\label{fig:Graficasai1LVai}Behavior of one of the variable pairs
considered for the comparison, showing the jitter of the processing
period running continuously for 24 hours.}
\end{figure}

Figure \ref{fig:Graficasai1LVai} displays the behavior of two process
variables of the type of analog input during a time interval of 24
hours of continuous operation. Each of the two variables corresponds
to one of the solutions implemented in this work: the solution based
on the use of LabVIEW and the one based on the standard EPICS system.
The figures shown the distribution of the processing period around
the theoretical period of 1 second for 24 hours .The initial analysis
of the obtained data, which is shown in Figure \ref{fig:Graficasai1LVai},
indicate very good performance with a similar jitter in both cases.
Moreover, the rest of the signals involved in the comparison experiment
behave in a similar way. In the present status of the study the tests
have been running continuously for several days, however, a longer
time interval is required to fully ensure the adequate performance
of the proposed approach for the fast controller integration and to
provide more reliable comparison results.

The present test bench is also used for the validation of the archiving
system. By using the Hyperarchiver archiving tool, it can be analyzed
the time span between consecutive timestamps. This data can be later
used to analyze the quality of the archiving process. Thus, the test
bench provides an appropriate mechanism for the improvement and adjustment
of the parameters of the archiving system, which eventually might
help avoiding the data loss.

\subsection*{Worst case test results}

The main objective of the proposed fast controller is oriented to
the systems which require a high performance with a high speed data
acquisition and time synchronization. It is not oriented to applications
aimed to a high data volume processing. However, in order to test
the behavior of the system in different scenarios, the next situations
have been considered: 
\begin{itemize}
\item An increase in the number of access in the EPICS network. The increment
of \emph{get} petitions does not lead to any problem when packets
up to 180 PVs are requested at once.
\item An increase in the number of records until 16000 in each controller,
i.e., the number of PVs is approximately 8 times higher. The results
are similar in both cases, showing a good behavior. However, the deployment
of the application, that is, the initialization time, is higher using
the proposed solution than the standard solution. 
\item An increase in the number of EPICS monitors. When the number of synchronous
EPICS monitors exceeds 80 monitors, the LabVIEW based system presents
robustness problems, the controller becomes unstable and randomly
hangs on data requests, issue not observed in the standard solution. 
\end{itemize}
The results obtained show that the standard solution is more robust
in applications with a large number of PVs. This fact is not a surprise
and this configuration is preferable to be used when a large number
of PVs is required, including simultaneous monitorization. In any
case, this issue is not an important drawback of the proposed solution
since it is oriented to control applications with high processing
requirements but not with high data volume. Moreover, the election
in the tests of the monitors has been done in the worst conditions
to detect the limit of the system. In this implementation, the simultaneous
monitors get the data directly from the LabVIEW EPICS server (using
the DSC module). Nevertheless, the monitors can be redesigned to minimize
the problems detected using the LabVIEW/EPICS interface.

\section{Conclusions}

In this work a methodology for the design of controllers providing
high performance is described. The system is very versatile and powerful
thank to the combination of the use of FPGAs, real time controllers
and high performance data acquisition cards. In particular, the methodology
is oriented to the field of control systems in particle accelerators,
related with the fast controllers, and facilitates the integration
in EPICS networks. The main advantages of the proposed methodology
are, on one hand, the use of fast, flexible and versatile devices
and, on the other hand, those devices are easily obtained, since they
are COTS and general purpose hardware. Moreover, an integration methodology
in EPICS networks is proposed, which it is transparent from the point
of view of this middleware. This integration facilitates the communication
over EPICS with other controllers.

The architecture details differ from the standard EPICS approach in
similar facilities. In the present work it is used a LabVIEW RT based
fast controller in addition to an EPICS server for its integration
in the general control network. This is one of the main contributions
of the present work, that is, the design and implementation of a communication
gateway between EPICS and LabVIEW providing a full EPICS IOC. This
approach has several advantages derived from an easier and agile implementation
thanks to the use of commercial drivers. On the contrary, several
potential drawbacks arise, being the most import one the reliability,
which must be assured for those schemes, since its usage in other
facilities is not extended. In addition, the control of the internal
details of the software relays on external companies.

In this work, a laboratory test bench for the comparison of the proposed
methodology with standard EPICS architectures is presented. The results
obtained in the experimental comparison are positive, since the proposed
methodology gives similar results comparing to a standard solution
while the development process is faster. In any case, the results
show a reliable system, comparable with other solutions used nowadays,
very important characteristic in a real facility. However, the worst
case tests show the limitation of the LabVIEW EPICS server when using
simultaneous monitors. On this sense, the proposed gateway can used
to improve this issue, adding a new functionality to the presented
methodology.

The conclusions obtained from this work encourage to follow working
on this idea, performing more detailed tests in order to assure the
required performance of the proposed tool and its advantages. In particular,
longer experiments are necessary, including a more diverse set of
available control actions. On the other hand, the analysis of the
real-time performance integrated in the EPICS network is useful, which
could allow the analysis of the fulfillment of time restrictions from
the point of view of the local controller and the EPICS network, simultaneously.

\section{Acknowledgments}

The authors are very grateful by the partial support of this work
to the Basque Govern by mean of the project \emph{Grupos de Investigación
GIU06/04} and to the Consorcio ESS Bilbao.

\bibliographystyle{IEEEtran}
\bibliography{biblio_RevSciInstr}

\end{document}